\documentclass[usenatbib]{mn2e}

\usepackage{amsmath}
\usepackage{graphicx}
\usepackage{txfonts}
\usepackage{enumerate}
\usepackage{bm}
\usepackage{natbib}
\usepackage{listings}           

\usepackage{hyperref}
\hypersetup{colorlinks=true,
            citecolor=blue,
            linkcolor=blue}

\def\be {\begin{equation}}
\def\ee {\end{equation}}

\def\tt {\{t\}}

\def\nul#1{}

\renewcommand*{\vec}[1]     {\boldsymbol{#1}}

\begin{document}
%
\title[Reversible time-stepping]{Reversible time-step adaptation for the integration of few-body systems}

\author[Boekholt, Vaillant and Correia]{Tjarda C. N. Boekholt$^{1}$\thanks{E-mail: tjarda.boekholt@physics.ox.ac.uk}, Timoth\'ee Vaillant$^{2}$ and Alexandre C. M. Correia$^{2,3}$\\
$^{1}$Rudolf Peierls Centre for Theoretical Physics, Clarendon Laboratory, Parks Road, Oxford, OX1 3PU, United Kingdom\\
$^2$ CFisUC, Departamento de F\'isica, Universidade de Coimbra, 3004-516 Coimbra, Portugal\\
$^3$ IMCCE, UMR8028 CNRS, Observatoire de Paris, PSL Universit\'e, 77 Av. Denfert-Rochereau, 75014 Paris, France\\
}

\date{\today}

\maketitle

\begin{abstract}
   The time step criterion plays a crucial role in direct $N$-body codes. If not chosen carefully, it will cause a secular drift in the energy error. Shared, adaptive time step criteria commonly adopt the minimum pairwise time step, which suffers from discontinuities in the time evolution of the time step. This has a large impact on the functioning of time step symmetrisation algorithms. We provide new demonstrations of previous findings that a smooth and weighted average over all pairwise time steps in the N-body system, improves the level of energy conservation. Furthermore, we compare the performance of 27 different time step criteria, by considering 3 methods for weighting time steps and 9 symmetrisation methods. We present performance tests for strongly chaotic few-body systems, including unstable triples, giant planets in a resonant chain, and the current Solar System. We find that the harmonic symmetrisation methods  (methods A3 and B3 in our notation) are the most robust, in the sense that the symmetrised time step remains close to the time step function. Furthermore, based on our Solar System experiment, we find that our new weighting method based on direct pair-wise averaging (method W2 in our notation), is slightly preferred over the other methods. 
\end{abstract}



\section{Introduction}\label{sec:introduction}

$N$-body codes consist of two main ingredients: the integration method and the time step criterion. 
Our main focus is to implement and compare various time step criteria, and to measure their effectiveness in conserving energy for a wide range of initial conditions.   

The simplest time step criterion is the constant one. This criterion is commonly implemented in $N$-body codes for planetary systems, where orbital periods are not expected to change much. Usually, the time step is chosen as some fraction of the shortest, initial, orbital period in the system \citep[e.g.][]{Quinn_1991, Jones_2001, Ito_2002, Bolmont_2015}. 

If instabilities grow during the evolution, potentially resulting in planet-planet scattering \citep{Chambers_1996}, then the time step size has to be reduced in order to resolve the dynamics accurately \citep{Chambers_1999, Petit_2019, Rein_Hernandez_2019}. Varying the time step is very common in $N$-body codes for dense stellar systems. Here, stellar orbits are chaotic, and close binaries are formed and destroyed continuously \citep[e.g.][]{Heggie75}. It is then much preferred to implement a time step adaptation method. There are various ways to do this, including: 1) assign each pair of bodies in the current configuration, a time step according to some criterion, and then adopt the minimal value, or 2) evaluate each body individually, and determine on what time scale its orbit is changing, and then adopt the minimal value over all bodies. Commonly used examples of the first case are the pairwise free-fall and flyby time scales \citep[e.g.][]{Hut_2007, Pelupessy_Janes_2012}, and for the second case we mention the criterion by Sverre Aarseth based on the acceleration and its derivatives \citep{Aarseth_2003}.  

Shared time step schemes, where each body adopts the same time step size, are simple to implement, but too expensive to be used for large-N systems \citep{Makino_1991, Trenti_Hut_2008}. For example, a star in the cluster's halo should not have to suffer the same small time step as two stars in a close encounter in the core. State of the art $N$-body codes for large stellar systems therefore employ an individual time step criterion \citep[e.g.][]{Wang_2020}. Synchronisation and predictions of positions and velocities play an important role, which is greatly facilitated by using block time steps \citep[e.g.][]{Makino_Hut_2006}. 

If the $N$-body system under study allows for a constant time step criterion, then this would be greatly preferred due to its simplicity, numerical efficiency and its compatibility with symplectic integrators.
If an adaptive scheme is required, this not only comes with an extra cost of evaluating the time step, but the favourable symplectic properties also tend to be lost \citep[e.g.][]{Dehnen_2017}. 

Much research has been put into making adaptive time step schemes compatible with symplectic integration methods. For an extensive review, we refer the reader to \cite{Dehnen_2017}. The main idea however, is obtained by considering an analogy to Noether's theorem on time invariance and energy conservation. If the level of time-reversibility of an $N$-body integration is increased, then correspondingly, it might be expected that the level of energy conservation increases.
This idea led to the concept of time step symmetrisation methods \citep{Quinlan_Tremaine_1990, Hut_Makino_1995, Dehnen_2017}. As a concrete example, we consider a body on an eccentric orbit somewhere between pericenter and apocenter. Based on its current position and velocity, we would calculate a time step size. This step size would be the same irrespective of whether the body was approaching pericenter or apocenter, e.g. the direction of the orbit. If we consider having taken an integration step towards pericenter, then the newly evaluated time step size will be shorter, as we are closer to pericenter. However, if we would take this new time step, but integrate back along the orbit, we would not end up on our initial condition. 
In other words, adaptive time step functions based on the current snapshot only, do not result in time-symmetric integrations. As a consequence, the lack of error cancellation results in a secular drift in the energy error.  
In order to achieve a time-symmetric integration, 
we require a symmetrised time step, $h$, which depends symmetrically on the old and new states.

However, as discussed by \citet{Hands19}, an adaptive and symmetrised time step is generally not sufficient. Consider a closest pair of bodies with a certain time step, $T_0$. We can imagine this closest pair to be receding from each other, such that $T_0$ increases. At the same time, there is another pair whose time step is $T_1 > T_0$. However, this pair is approaching, and therefore $T_1$ is decreasing. There will be a transition point at which $T_0 = T_1$, but each pair will have a different value for the derivative. In short, if there is a transition to a new pair of bodies with the smallest time step, a discontinuity can arise in the time step function and its derivative. 
The discontinuities arise from adopting the minimal time step over all pairs of bodies. Instead, \citet{Hands19} propose to implement a weighted average, such that small time step values have a larger weight. Such a scheme makes use of all the orbital frequencies in the $N$-body system instead of just one. 

Our aim is to expand on the work by \cite{Hands19}, and to test for the positive effect of smoothness for a variety of time step criteria and N-body systems. We implement three different methods for weighting time steps, including the algorithm by \cite{Hands19} and two of our own. We will also compare nine different time step symmetrisation methods, most of them based on previous literature, but complemented with three new ones. This results in a total of 27 different combinations. Furthermore, we will apply these time stepping methods to different types of astrophysical systems. This allows us to determine if there is a single, optimal time step criterion for chaotic few-body systems.

\section{Methods}\label{sec:methods}

We start by describing two fourth-order integration methods adopted for our simulations. Then we present the three ingredients for the time step criteria: 1) a fixed pairwise time step function, 2) three weighting methods for calculating a smooth and ``global" time step, and 3) nine different time step symmetrisation methods. An overview of these ingredients is provided in Tab.~\ref{tab:dt}. 

\subsection{Fourth-order integrators}

As the benchmark integrator, we will adopt the fourth-order scheme by \citet{McLachlan_1995} denoted by (order 4, S, $m=5$). This method is time-reversible and symplectic, when combined with a constant time step. 
The integration map consists of a symmetric sequence of kicks and drifts with a total of 5 calculations of the acceleration, as follows:

\begin{equation}
    \begin{split}
    \phi_{\rm{MCL4}}\left( h \right) = K\left( a_1 h \right) D\left( b_1 h \right) K\left( a_2 h \right) D\left( b_2 h \right) K\left( a_3 h \right) D\left( b_3 h \right) \\ 
    K\left( a_3 h \right) D\left( b_2 h \right) K\left( a_2 h \right) D\left( b_1 h \right) K\left( a_1 h \right).        
    \end{split}
\end{equation}

\noindent Here $h$ is the time step, $K$ and $D$ refer to the usual Kick and Drift operators (as used in the Verlet-Leapfrog integrator), and the coefficients are given by $a_1 = \left(14-\sqrt{19}\right)/108$, $a_2 = \left( 20-7\sqrt{19} \right)/108$, $a_3 = \left( 1 - 2a_1 - 2a_2 \right)/2$, $b_1 = 2/5$, $b_2 = -1/10$, and $b_3 = 1-2b_1-2b_2$.
This sequence also includes negative time steps, but \citet{McLachlan_1995} demonstrates that this integrator is among the most effective fourth-order integrators constructed by a composition of kicks and drifts.

In order to test for the generality of a potential optimal time step criterion, we implement a second integrator. We adopt the method by \citet{chinchen05} (following a suggestion by Walter Dehnen), which is also a fourth-order method, but which only includes positive time steps. Although this requires the calculation of the force gradient, the acceleration only needs to be calculated twice per step (together with an initial calculation at the start of the simulation). The integration map is given by:

\begin{equation}
    \begin{split}
    \phi_{\rm{CC4}}\left( h \right) = K\left( u_1 h \right) D\left( u_2 h \right) \tilde{K}\left( u_3 h \right) D\left( u_2 h \right) K\left( u_1 h \right).        
    \end{split}
\end{equation}

\noindent The operator $\tilde{K}$ represents the kick with the adjusted acceleration taking into account the force gradient \citep{chinchen05, Dehnen+17}. The coefficients are given by $u_1 = 1/6$ and $u_2 = 1/2$ and $u_3 = 2/3$. We will abbreviate the names of the two integrators as MCL4 \citep{McLachlan_1995} and CC4 \citep{chinchen05}.

\subsection{Time step function}\label{sec:dts}

For the time step function, $T$, we adopt the commonly used combination of the pairwise free-fall time scale and the pairwise flyby time scale \citep[e.g.][]{Pelupessy_Janes_2012, Hands19}, respectively given by:

\begin{equation}
    T_{ij,0}^{-4} = \frac{\mu_{ij}^2}{r_{ij}^6}, 
    \label{eq:Tij0}
\end{equation}

\noindent and 

\begin{equation}
    T_{ij,1}^{-2} = \frac{v_{ij}^2}{r_{ij}^2}. 
    \label{eq:Tij1}
\end{equation}

\noindent Here, $\vec{r}_{ij} = \vec{r}_i - \vec{r}_j$ is the separation between bodies $i$ and $j$, $\vec{v}_{ij} = \vec{v}_i - \vec{v}_j$ is the relative velocity, and $\mu_{ij} = G\left( m_i + m_j \right)$ is the gravitational parameter. 
The free-fall time scale can also be interpreted as the separation divided by the circular velocity corresponding to that separation. The flyby time scale is required when the relative speeds much exceed this circular velocity.

The derivatives of the time step functions are required for some time step symmetrisation methods. They are given by

\begin{equation}
    T_{ij,0}^{-1} \dot{T}_{ij,0} = \frac{3}{2} \frac{\vec{r}_{ij} \cdot \vec{v}_{ij}}{r_{ij}^2},
\end{equation}
\begin{equation}
    T_{ij,1}^{-1} \dot{T}_{ij,1} = \frac{\vec{r}_{ij} \cdot \vec{v}_{ij}}{r_{ij}^2} - \frac{\vec{v}_{ij} \cdot \vec{a}_{ij}}{v_{ij}^2},
\end{equation}

\noindent The derivative of the flyby time scale includes the relative acceleration between a pair of bodies. We wish to emphasise here that this should be the difference in the total acceleration of the two bodies, i.e. $\vec{a}_{ij} = \vec{a}_i - \vec{a}_j$. If on the other hand, the approximation is made of only evaluating their mutual acceleration, which depending on the integration method is more convenient to calculate, then the derivative of the time step is not guaranteed to be accurate. We find that this can cause time step symmetrisation methods to fail, resulting in excessively small or large time steps.

Other time step functions could be defined, for example taking into account the accelerations. However, since some symmetrisation methods require the derivative of the time step, this would imply having to calculate the jerk (and potentially higher order derivatives), which leads to an increase of the computation time. We leave the performance comparisons with such time step criteria for elsewhere.

A common method then for calculating the next integration time step is to evaluate the free-fall and flyby time scale for all pairs of bodies in the system, and then to adopt the minimal value. This ensures the highest frequency encounters are resolved. However, this also results in discontinuities in the evolution of the time step function. An improved method is to consider a smooth combination of all pairwise time steps by introducing weights.

\subsection{Weighting time steps}\label{sec:weights}

A general method for combining multiple time steps into a single value is the following:

\begin{equation}
    T = \left( \frac{\sum_{k} T_{k}^{m-n} }{\sum_{k} T_{k}^{-n} } \right)^{\frac{1}{m}},
    \label{eq:T}
\end{equation}

\noindent with $m>0$, $n>0$ and the summation is over all the time step values under consideration (pairwise or per body as we will discuss later in this section). For example, we can set $m=1$ resulting in

\begin{equation}
    T = \frac{\sum_{k} T_{k}^{-n} T_k }{\sum_{k} T_{k}^{-n} } \equiv \frac{\sum_{k} w_k T_k }{\sum_{k} w_k }, 
    \label{eq:T_w}
\end{equation}

\noindent where we defined the time step weight $w_k = T_k^{-n}$. It is crucial for this weighted time step to remain close to the minimal pairwise time step in order to resolve close encounters. Another combination is ($m=2$, $n=10$) \citep{Hands19}, which gives

\begin{equation}
    T = \left( \frac{\sum_{k} T_{k}^{-8} }{\sum_{k} T_{k}^{-10} } \right)^{\frac{1}{2}}.
    \label{eq:T_482}
\end{equation}

\noindent Alternatively, for $m=n$ we obtain 

\begin{equation}
    T = \left( \frac{\sum_{k} T_{k}^{-n}}{N_T} \right)^{-\frac{1}{n}},      
    \label{eq:T_q}
\end{equation}

\noindent with $N_T$ the number of time steps to be summed. As described by \cite{Hands19}, this combination has a problematic feature. In the limit that all time steps are similar, i.e. $T_k \approx \tau$, we obtain 

\begin{equation}
    T \approx \left( \frac{N_T \tau^{-n}}{N_T} \right)^{-\frac{1}{n}} = \tau. 
\end{equation}

\noindent However, if there is a single time step which is by far the smallest, $T_{\rm{min}}$, then we obtain

\begin{equation}
    T \approx \left( \frac{T_{\rm{min}}^{-n}}{N_T} \right)^{-\frac{1}{n}} = T_{\rm{min}} N_T^{\frac{1}{n}}. 
\end{equation}

\noindent Hence, for large-$N$ systems, the time step $T$ becomes much larger than $T_{\rm{min}}$, resulting in large discretisation errors. A slightly improved method is to multiply $T$ by $N_T^{-\frac{1}{n}}$ in Eq.~\eqref{eq:T_q}, resulting in

\begin{equation}
    T = \left( \sum_{k} T_{k}^{-n} \right)^{-\frac{1}{n}}.
    \label{eq:Tmeqn}
\end{equation}

\noindent In the regime of a single smallest time step, this reduces to 

\begin{equation}
    T \approx \left( T_{\rm{min}}^{-n} \right)^{-\frac{1}{n}} = T_{\rm{min}},
\end{equation}

\noindent while for the similar time steps case we obtain

\begin{equation}
    T \approx \left( N_T \tau^{-n} \right)^{-\frac{1}{n}} = \tau N_T^{-\frac{1}{n}}.
\end{equation}

\noindent Hence, in this case the time step $T$ becomes increasingly smaller than necessary for larger $N$ systems. We note that these dependencies on $N_T$ cancel out for combinations where $m \neq n$ in Eq.~\eqref{eq:T}. 

On the other hand, in the case where $m \neq n$, problems might arise with keeping the smoothed time step close to the minimum time step, for an increasing number of bodies. This can be more clearly seen by rewriting Eq.~\eqref{eq:T_w} as follows:

\begin{equation}
    \frac{T}{T_{\rm{min}}} = \frac{\sum_{k} T_{k}^{1-n}}{T_{\rm{min}} \sum_{k}  T_{k}^{-n} } = \frac{1 + \sum_{k \neq l} \left(\frac{T_k}{T_{\rm{min}}}\right)^{1-n}}{1 + \sum_{k \neq l}  \left( \frac{T_k}{T_{\rm{min}}}\right)^{-n} } < f_T,
    \label{eq:logN}
\end{equation}

\noindent where we divided by the minimum pairwise time step, took the minimum time step outside the summation, and where $T_l = T_{\rm{min}}$. Here, $f_T$ gives the maximum value of the fraction $T/T_{\rm{min}}$.  
We consider a certain number of bodies $N=N_0$ and a weight parameter $n=n_0$, such that Eq.~\eqref{eq:logN} holds for a given value of $f_T$. If we now increase the number of bodies to $N=N_1 > N_0$, while assuming that the distribution of time steps is fixed, then this amounts to multiplying the summations by a factor of about $c=\left(\frac{N_1}{N_0}\right)^\alpha$, with $\alpha = 1$ or $2$ depending on whether the sum is over all bodies or over all pairs. This might result in a violation of the constraint in Eq.~\eqref{eq:logN}. The increasing number of terms in the summation is to be compensated for by adjusting the weight parameter $n$ from $n_0$ to $n_1$. Considering a single term of the summation in the denominator, we require

\begin{equation}
    \left( \frac{T_k}{T_{\rm{min}}} \right)^{-n_0} \approx \left( \frac{N_1}{N_0} \right)^\alpha \left( \frac{T_k}{T_{\rm{min}}} \right)^{-n_1}, 
\end{equation}

\noindent which can be rewritten as

\begin{equation}
    n_1-n_0 \propto \log\left( \frac{N_1}{N_0} \right). 
\end{equation}

\noindent Hence, in order to keep the smoothed time step close to the minimum time step, we expect the weight parameter to scale with the number of bodies as $n \propto \log N$. We confirm this result in Appendix~\ref{app:logN}, and there we also demonstrate that for a particle number up to a few thousand, a weighting parameter of $n = 10$ is sufficient, consistent with the value adopted by \citet{Hands19}. We will adopt this value throughout our simulations, while also setting $m = 2$. Hence, we adopt the weighting method given by Eq.~\eqref{eq:T_482}. 

There are different ways to implement the summation, which can be over different types of time steps. In our first method, we will sum directly over all pairs of bodies, and add the free-fall and flyby time scales individually:

\begin{equation}
    T = \eta \left( \frac{\sum_{i<j} T_{ij,0}^{-8} + T_{ij,1}^{-8} }{\sum_{i<j} T_{ij,0}^{-10} + T_{ij,1}^{-10} } \right)^{\frac{1}{2}}
    \hspace{1cm}\left( \rm{method\,W1}\right).
    \label{eq:WIJ-1}
\end{equation}

\noindent Here we multiply the global time step, $T$, by the time step parameter, $\eta$, which allows the magnitude of the averaged time step size to be varied systematically. 
Alternatively, as proposed by \citet{Hands19}, we can first average the pairwise free-fall and flyby time scale according to

\begin{equation}
    T_{ij}^{-q} \equiv T_{ij,0}^{-q} + T_{ij,1}^{-q}.
    \label{eq:T01q}
\end{equation}

\noindent Following \citet{Hands19}, we adopt $q=4$, which gives a good balance between the amount of smoothing and computational efficiency (note that the time step functions were introduced in powers of 2 and 4). Our second method for weighting time steps is then given by:

\begin{equation}
    T = \eta \left( \frac{\sum_{i<j} T_{ij}^{-8} }{\sum_{i<j} T_{ij}^{-10} } \right)^{\frac{1}{2}}
    \hspace{1cm}\left( \rm{method\,W2}\right).
    \label{eq:WIJ-2}
\end{equation}

\noindent with $T_{ij}$ given by Eq.~\eqref{eq:T01q} with $q=4$. Our third weighting method follows the method by \citet{Hands19}, which introduces another level of averaging. After combining the free-fall and flyby time scales as in our second method, we first calculate averaged time steps per body according to Eq.~\eqref{eq:Tmeqn}: 

\begin{equation}
    T_{i}^{-n} \equiv \sum_{j \neq i} T_{ij}^{-n}.
    \label{eq:per_body}
\end{equation}

\noindent Although the value of $n$ can be different than the value of $q$ in Eq.~\eqref{eq:T01q}, we follow the prescription by \citet{Hands19} by setting $q=n=4$. Subsequently, the ``per body'' time steps are averaged similar to the other methods:

\begin{equation}
    T = \eta \left( \frac{\sum_{i} T_{i}^{-8} }{\sum_{i} T_{i}^{-10} } \right)^{\frac{1}{2}}
    \hspace{1cm}\left( \rm{method\,W3}\right).
    \label{eq:WI}
\end{equation}

\noindent Note that here the summation is over all bodies. We will abbreviate the three weighting methods as W1, W2 and W3, as noted behind their respective equations.
In our experiments, we will directly compare the performance of these three methods.

For completeness, we give the derivative of the generic expression for the summation of time steps (Eq.~\eqref{eq:T}): 

\begin{equation}
\begin{split}
    \dot{T} = & \frac{\eta}{m} \left( \frac{\sum_{k} T_{k}^{m-n} }{\sum_{k} T_{k}^{-n} } \right)^{\frac{1}{m}-1} \\
    & \left( \left(m-n\right) \frac{\sum_{k} T_{k}^{m-n-1} \dot{T}_k }{\sum_{k} T_{k}^{-n} } + n \frac{ \sum_{k} T_{k}^{m-n} \sum_{k} T_{k}^{-n-1} \dot{T}_k }{\left( \sum_{k} T_{k}^{-n} \right)^2} \right).
\end{split}
\label{eq:dT}
\end{equation}

\begin{table*}
    \centering
    \begin{tabular}{|l|l|l|l|l|}
    \hline
Abbreviation    &   Pairwise time step function &   Weighting method  &   Symmetrisation method   & Comment \\
    \hline
    
$T_{ij,0}$ &   Eq.~\eqref{eq:Tij0} &   & & Pairwise free fall time scale \\
$T_{ij,1}$ &   Eq.~\eqref{eq:Tij1} &   & & Pairwise flyby time scale \\
\hline
$T_{ij}$     &    $T_{ij,0}$, $T_{ij,1}$    & Eq.~\eqref{eq:T01q} with $q=4$ &  & Pairwise weighted time step \\
$T_{i}$     &     $T_{ij}$    & Eq.~\eqref{eq:per_body} with $n=4$ &  & Weighted time step per body \\
\hline
$W1$     &   $T_{ij,0}$, $T_{ij,1}$    &   Eq.~\eqref{eq:WIJ-1}  &  & Weighted time step over all pairs \\

$W2$     &   $T_{ij}$    &   Eq.~\eqref{eq:WIJ-2}  &  & Weighted time step over all pairs \\

$W3$     &   $T_{i}$    &   Eq.~\eqref{eq:WI}  &  & Weighted time step over all bodies \\
    \hline
A1 & & & Eq.~\eqref{eq:A1} & Linear \\
A2 & & & Eq.~\eqref{eq:A2} & Logarithmic \\
A3 & & & Eq.~\eqref{eq:A3} & Harmonic \\
B1 & & & Eq.~\eqref{eq:B1} & Linear, using $\tau=T$ and $\dot{T}$ \\
B2 & & & Eq.~\eqref{eq:B2} & Logarithmic, using $\tau=T$ and $\dot{T}$ \\
B3 & & & Eq.~\eqref{eq:B3} & Harmonic, using $\tau=T$ and $\dot{T}$ \\
C1 & & & Eq.~\eqref{eq:C1} & Linear, using $\tau = \left( h_{\rm{prev}} + h_{\rm{next}} \right)/2$ and $\dot{T}$\\
C2 & & & Eq.~\eqref{eq:C2} & Logarithmic, using $\tau = \left( h_{\rm{prev}} + h_{\rm{next}}\right)/2$ and $\dot{T}$\\
C3 & & & Eq.~\eqref{eq:C3} & Harmonic, using $\tau = \left( h_{\rm{prev}} + h_{\rm{next}}\right)/2$ and $\dot{T}$\\
    \hline
    \end{tabular}
    \caption{Overview of the time stepping methods used in this study. For methods $B$ and $C$, the quantity $\tau$ refers to the time scale for the change in $F\left( h \right)$. }
    \label{tab:dt}
\end{table*}

\subsection{Symmetrisation methods}

We distinguish between the symmetrised time step, $h$, and the time step function, $T$, both of which are functions of time, $t$. The time step function is calculated according to the algorithms described in the previous two subsections. The symmetrised time step is the one used for the actual integration. 
Explicit time step symmetrisation methods assume a function $F\left( T \right)$, and a symmetric combination of $F\left( h_{\rm{prev}} \right)$ and $F\left( h_{\rm{next}} \right)$, where $h_{\rm{prev}}$ and $h_{\rm{next}}$ are the previous and next symmetrised time step, respectively. 
Inspired by the overview of symmetrisation methods by \citet{Dehnen_2017}, we adopt the following general approaches for reversible time step adaptation:

\begin{flalign}
\; & (A) & F(T) &= \frac{F(h_{\rm{prev}}) + F(h_{\rm{next}})}{2} & & \rm{\left(flip\right),}\\
   & (B) & \frac{dF}{dt}\left( T \right) &= \frac{ F\left(h_{\rm{next}}\right) - F\left(h_{\rm{prev}}\right)}{T} & & \rm{\left(integration\right),}\\
   & (C) & \frac{dF}{dt}\left(T\right) &= 2 \frac{ F\left(h_{\rm{next}}\right) - F\left(h_{\rm{prev}}\right) }{h_{\rm{next}} + h_{\rm{prev}}} & & \rm{\left(improved\:integration\right)}. 
\end{flalign}

\noindent Note that for method $B$, the right-hand side of the equation considers the change in $F$ over a time scale $T$, while method $C$ considers a time scale $\left( h_{\rm{next}} + h_{\rm{prev}} \right)/2$. Method $C $ (proposed by Walter Dehnen (personal communication)) is an improvement over method $B$, because integration errors can result in $T$ diverging from $h$.
For the function $F$, we adopt the following three functions:

\begin{flalign}
\; & (1) & F\left( T \right) &= T & & \rm{\left(Linear\right)}, \\
\; & (2) & F\left( T \right) &= \ln{T} & & \rm{\left(Logarithmic\right)}, \\
\; & (3) & F\left( T \right) &= \frac{1}{T} & & \rm{\left(Harmonic\right)}.
\end{flalign}

\noindent Different symmetrisation methods are obtained by considering all the permutations of the symmetrisation methods ($A, B, C$) and functions ($1, 2, 3$).

The first three symmetrisation methods are derived from method $A$ \citep[see][]{huang97, HOLDER2001367, leimkuhler_reich_2005}:

\begin{flalign}
   & & h_{\rm{next}} &= 2T - h_{\rm{prev}} &\rm{(A1)}
    \label{eq:A1} \\    
   & & h_{\rm{next}} &= \frac{T^2}{h_{\rm{prev}}} &\rm{(A2)}
    \label{eq:A2} \\
   & & h_{\rm{next}} &= \frac{1}{\frac{2}{T} - \frac{1}{h_{\rm{prev}}}} &\rm{(A3)}
    \label{eq:A3}
\end{flalign}

\noindent A second family of symmetrisation methods is derived from method $B$:

\begin{flalign}
   & & h_{\rm{next}} &= h_{\rm{prev}} + \dot{T}T &\rm{(B1)}
    \label{eq:B1} \\    
   & & h_{\rm{next}} &= h_{\rm{prev}} e^{\dot{T}} &\rm{(B2)}
    \label{eq:B2} \\
   & & h_{\rm{next}} &= \frac{1}{\frac{1}{h_{\rm{prev}}} - \frac{\dot{T}}{T}} &\rm{(B3)}
    \label{eq:B3}
\end{flalign}

\noindent These methods were introduced by \citet[][B2]{Dehnen_2017} and \citet[][B3]{Hairer_Soderlind_2005}, and also require the derivative of the smoothed global time step (Eq.~\eqref{eq:dT}). 
A third family of symmetrisation methods is derived from method $C$ \citep[see][]{Dehnen_2017, hairer06}:

\begin{equation}
    \dot{T} = 2 \frac{h_{\rm{next}}-h_{\rm{prev}}}{h_{\rm{next}}+h_{\rm{prev}}}
    \label{eq:C2_lin}
\end{equation}
\begin{equation}
    \frac{1}{T} \dot{T} = 2 \frac{\ln{h_{\rm{next}}}-\ln{h_{\rm{prev}}}}{h_{\rm{next}} + h_{\rm{prev}}}
    \label{eq:C2_0}
\end{equation}
\begin{equation}
    -\frac{1}{T^2} \dot{T} = 2 \frac{\frac{1}{h_{\rm{next}}}-\frac{1}{h_{\rm{prev}}}}{h_{\rm{next}} + h_{\rm{prev}}}
    \label{eq:C2_har}
\end{equation}

\noindent Here, the linear case (Eq~\eqref{eq:C2_lin}) can easily be rewritten in terms of $h_{\rm{next}}$. The harmonic case (Eq.~\eqref{eq:C2_har}) can be rewritten as a second-order polynomial in $h_{\rm{next}}$. Solving this results in two possible values for $h_{\rm{next}}$, both of which need to be calculated and compared as $\dot{T}$ can vary in sign as time evolves. The logarithmic case (Eq.~\eqref{eq:C2_0}) can only be solved using some iterative root-finding algorithm, which tend to be expensive. For this particular case, we will instead consider the following relation:

\begin{equation}
    \frac{1}{T}\dot{T} = \frac{1}{h} \dot{h} = \frac{2}{h_{\rm{next}}+h_{\rm{prev}}}\frac{h_{\rm{next}}-h_{\rm{prev}}}{\frac{h_{\rm{next}}+h_{\rm{prev}}}{2}} = 4 \frac{h_{\rm{next}}-h_{\rm{prev}}}{\left( h_{\rm{next}} + h_{\rm{prev}} \right)^2}
\end{equation}

\noindent This relation can also be rewritten as a second-order polynomial in $h_{\rm{next}}$, and we will use this relation instead of Eq.~\eqref{eq:C2_0}. The third family of symmetrisation methods is then given by:

\begin{equation}
    h_{\rm{next}} = h_{\rm{prev}} \frac{2 + \dot{T}}{2 - \dot{T}} \hspace{1cm} \rm{(C1)}
    \label{eq:C1}
\end{equation}
\begin{equation}
    h_{\rm{next}} = 
    \begin{cases}
    h_{\rm{prev}}, & \text{if $\xi = 0$ or $D_2 < 0$}.\\
    -h_{\rm{prev}} + \frac{2}{\xi} \left( 1 + \sigma \sqrt{D_2} \right) & \text{otherwise}.
    \end{cases}
    \hspace{0.5cm} \rm{(C2)}
    \label{eq:C2}
\end{equation}
\begin{equation}
    h_{\rm{next}} = 
    \begin{cases}
    h_{\rm{prev}}, & \text{if $\gamma = 0$ or $D_3 < 0$}.\\
    -\frac{h_{\rm{prev}}}{2} - \frac{1}{2 \gamma h_{\rm{prev}}} \left( 1 + \sigma \sqrt{D_3} \right) & \text{otherwise}.
    \end{cases}
    \hspace{0.5cm} \rm{(C3)}
    \label{eq:C3}
\end{equation}

\noindent where we defined $\xi \equiv \frac{\dot{T}}{T}$, $\gamma = -\frac{\dot{T}}{2 T^2}$, $D_2 = 1-2\xi h_{\rm{prev}}$ and $D_3 = \gamma^2 h_{\rm{prev}}^4 + 6\gamma h_{\rm{prev}}^2 + 1 $. For Eqs.~\eqref{eq:C2} and \eqref{eq:C3}, we have two solutions for $h_{\rm{next}}$, one for $\sigma=+1$,
the other for $\sigma=-1$. When the two solutions have a different sign, we choose
the positive solution, when the solutions have the same sign, we choose the one
with the minimum value of $|h_{\rm{next}}|$.

These explicit symmetrisation methods are not self-starting and an initial value for $h_{\rm{prev}}$ has to be determined. In Appendix~\ref{app:hprev}, we demonstrate that for an eccentric binary system, the following expression is appropriate:

\begin{equation}
h_{\rm{prev,0}} = \frac{T}{1 + \frac{1}{2}\dot{T}}.     
\end{equation}

\noindent We will adopt this initial value throughout our simulations. Other initial values can be obtained based on alternative symmetrisation methods. However, the effect of a slightly different initial value is expected to be marginal as it only concerns the initialisation.

\begin{figure*}
\centering
\begin{tabular}{c}
\includegraphics[height=0.377\textwidth,width=0.98\textwidth]{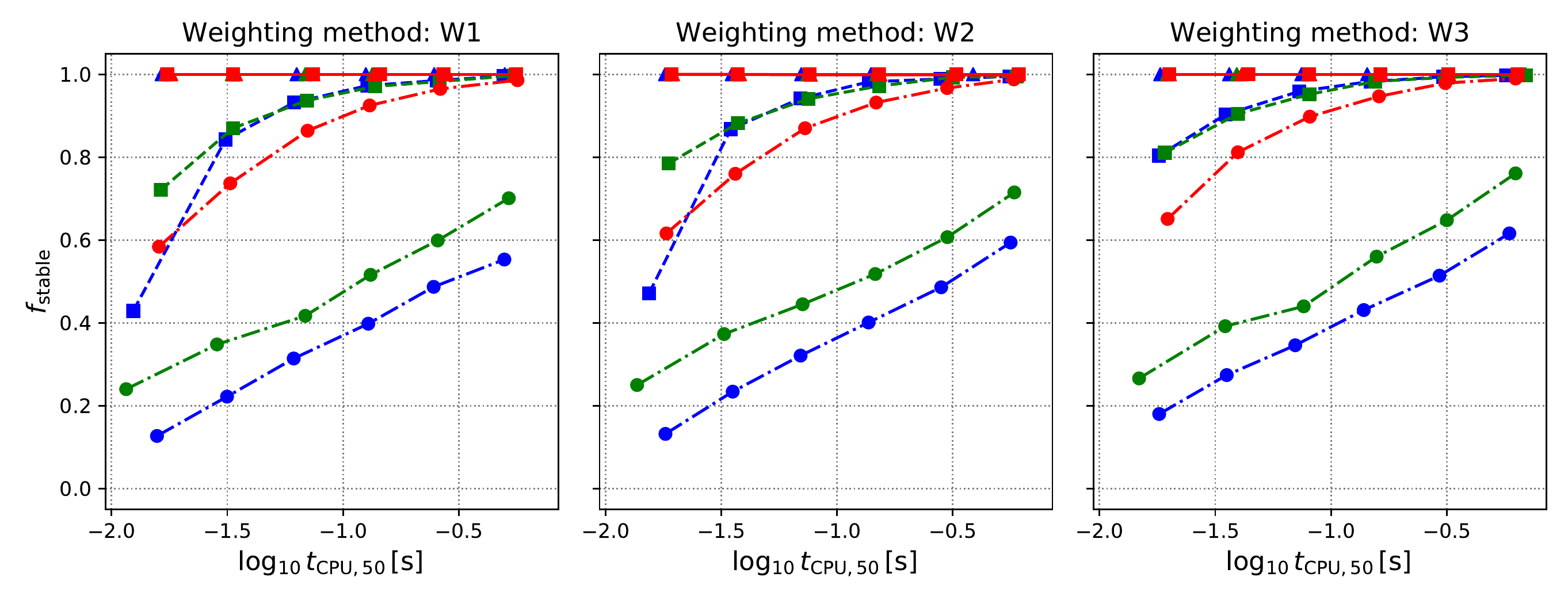} \\
\includegraphics[height=0.377\textwidth,width=0.98\textwidth]{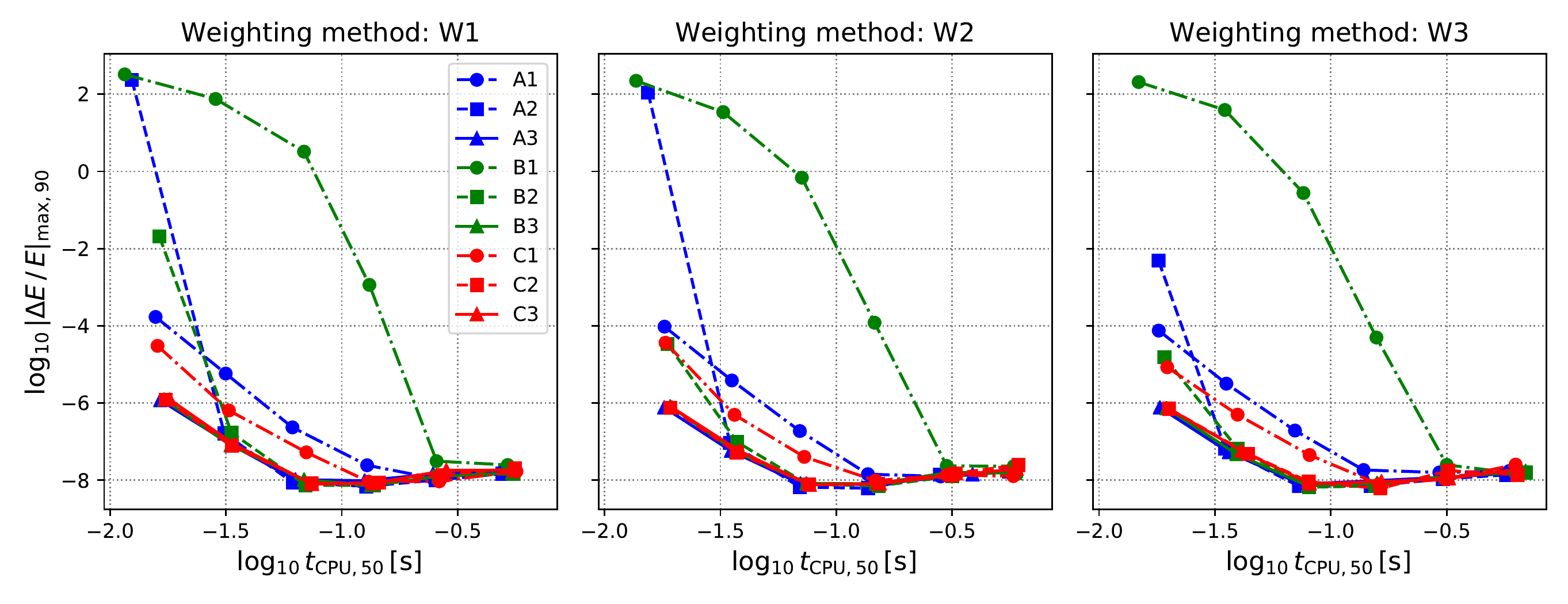} \\
\end{tabular}  
\caption{ We generate a set of 1000 random, equal-mass triple systems drawn from a Plummer distribution. Ensembles of solutions are obtained by evolving the triples for 100 dynamical times, using different time step weighting methods, symmetrisation methods, and values for the time step parameter. For each individual simulation, we measure the CPU running time ($t_{\rm{CPU}}$), the maximum, absolute value of the relative energy error ($\left| \Delta E\,/\,E \right|_{\rm{max}}$), and the maximum and minimum ratio of the symmetrised time step and the time step function ($f = h/T$). 
For each ensemble of simulations, 
we plot the $90^{th}$-percentile of the energy error ($\left| \Delta E\,/\,E \right|_{\rm{max,90}}$), and the fraction of ``stable'' simulations ($f_{\rm{stable}}$), for which $f$ remained within a factor 4 from $T$ at all times, with respect to the median CPU running time ($t_{\rm{CPU,50}}$). }
\label{fig:triple}
\end{figure*}

The symmetrisation methods defined above behave properly as long as the symmetrised time step, $h$, remains close to the time step function, $T$. During long term integrations of multi-body, chaotic systems however, this is not guaranteed. If $h \gg T$, energy errors will become intolerable, while if $h \ll T$, the simulation running time becomes excessive. 
As suggested by \citet{Hands19}, such instabilities can be contained by putting limits on the ratio of $h/T$. However, by resetting $h=T$ when a breach is detected, an irreversibility is introduced into the simulation. In order to test which time step criterion is the most robust, i.e. can keep $h$ close to $T$, we will implement rather mild limits in which $h$ can be a factor 128 larger or smaller than $T$. 
In practice, a good criterion will keep $h$ much closer to $T$. 
In this section, we have then defined three different methods for weighting time steps
and nine different time step symmetrisation methods ($\rm{Ai}$, $\rm{Bi}$, $\rm{Ci}$, $i=1,2,3$).
Therefore, there are 27 different possible combinations.
In Tab.~\ref{tab:dt}, we provide an overview of all time step ingredients used in this study.

\section{Results}

In this section, we will compare the performance of the 27 permutations of the time stepping methods described in Sec.~\ref{sec:methods}, for three different types of chaotic astrophysical systems. The first type consists of unstable triple stars, which will dissolve into a binary and single escaper. The second type is inspired by the Nice model, and considers unstable planetary systems. The time step size can vary drastically during planet-planet scattering. The third case considers a two million year integration of the current Solar System. By modelling each of these three systems, we can determine if there is a single, optimal time step criterion. 

\begin{figure*}
\centering
\begin{tabular}{c}
\includegraphics[height=0.377\textwidth,width=0.98\textwidth]{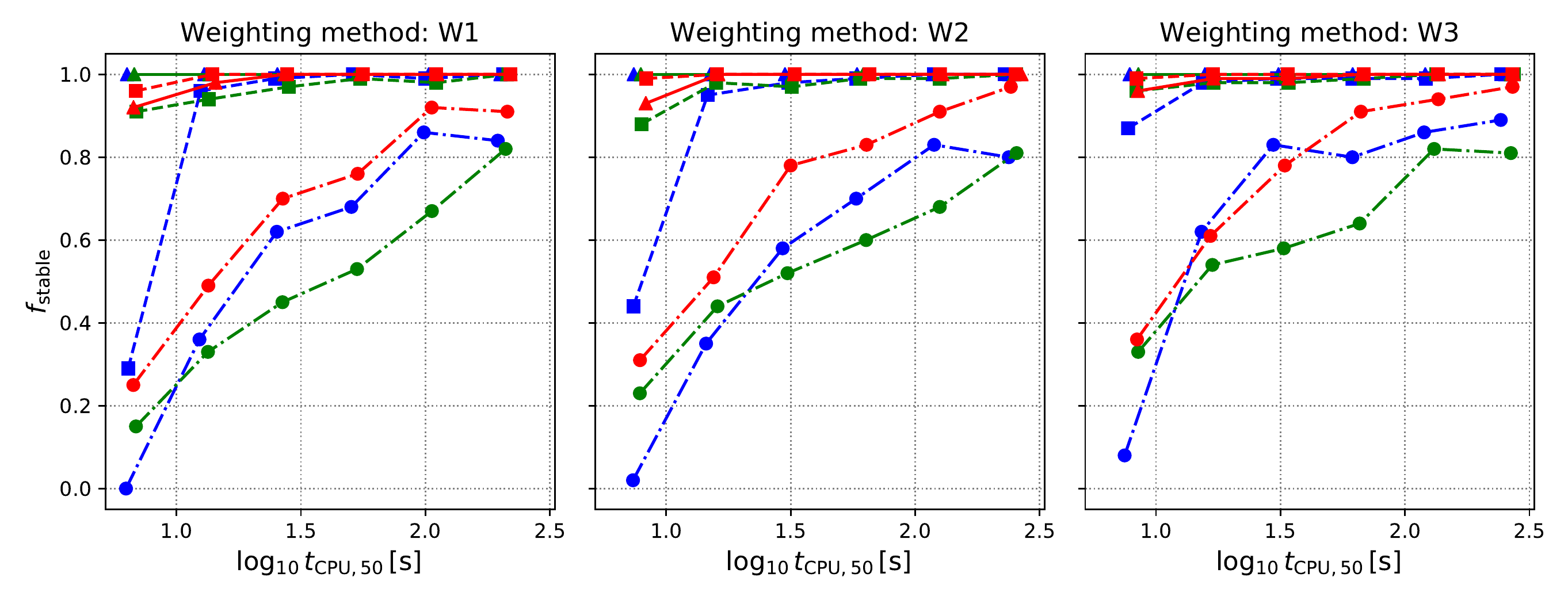} \\
\includegraphics[height=0.377\textwidth,width=0.98\textwidth]{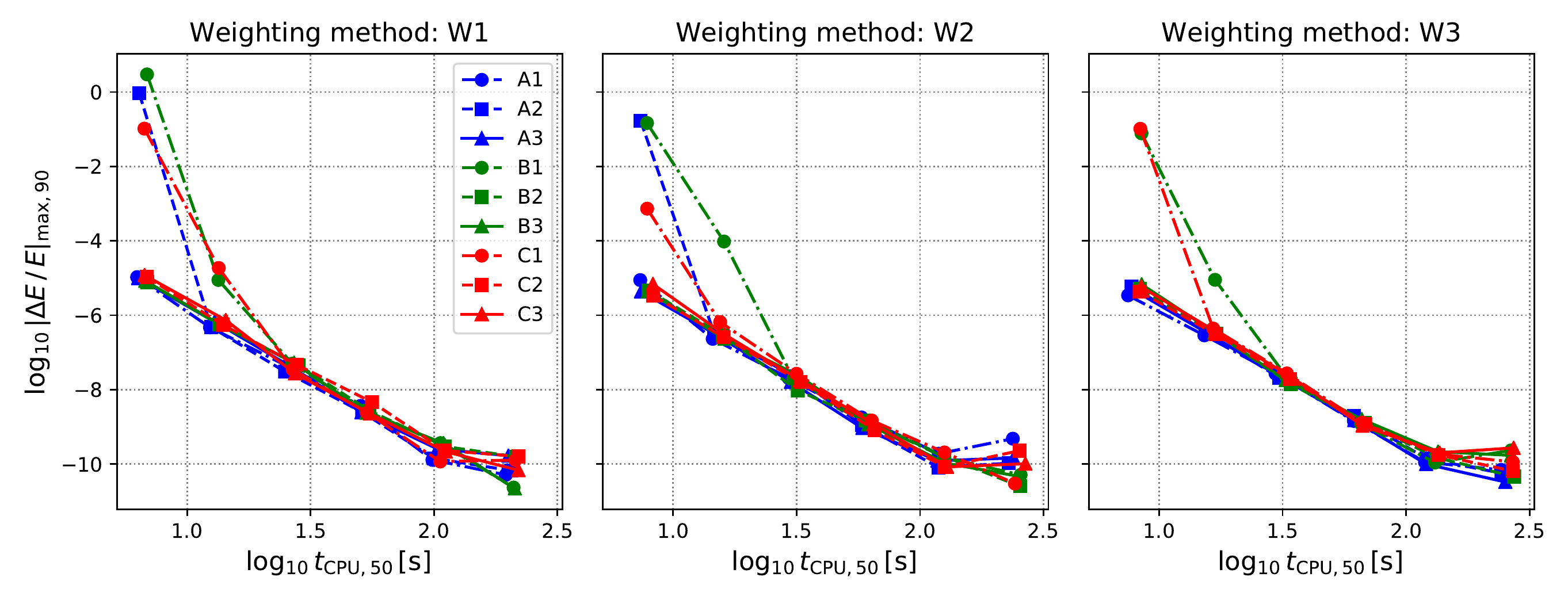} \\
\end{tabular}  
\caption{ We generate a set of 100 random initial realisation for the Nice model, including the Sun and the four giant planets. Ensembles of solutions are obtained by evolving the initial conditions for one million years, using different time step weighting methods, symmetrisation methods, and values for the time step parameter. 
For each individual simulation, we measure the CPU running time ($t_{\rm{CPU}}$), the maximum, absolute value of the relative energy error ($\left| \Delta E\,/\,E \right|_{\rm{max}}$), and the maximum and minimum ratio of the symmetrised time step and the time step function ($f = h/T$). 
For each ensemble of simulations, 
we plot the $90^{th}$-percentile of the energy error ($\left| \Delta E\,/\,E \right|_{\rm{max,90}}$), and the fraction of ``stable'' simulations ($f_{\rm{stable}}$), for which $f$ remained within a factor 4 from $T$ at all times, with respect to the median CPU running time ($t_{\rm{CPU,50}}$).  }
\label{fig:nice}
\end{figure*}

\subsection{Unstable triple systems}

Hierarchical triple star systems consist of an inner binary system, and a third star orbiting around the center of mass of the inner binary. If the orbits are sufficiently separated, the system is stable and shows interesting long term effects, such as Lidov-Kozai cycles \citep{Lidov_1962, Kozai_1962}. However, due to a combination of stellar evolutionary and dynamical effects, it is possible for the hierarchy to change. One scenario is that the inner binary orbit expands due to mass loss from stellar winds, thereby gradually reducing the hierarchy. Once the inner and outer orbits are sufficiently close according to a stability criterion for triples \citep[e.g.][]{2008LNP...760...59M}, then the triple has fully destabilised and will eventually break up \citep[for more details see][]{ST_TB_2021, Hamers_2021}. 

Due to the chaotic nature of triple systems, there is a sensitive dependence to small perturbations \citep[e.g.][]{TB_2020}. These could be caused by taking different time steps. It is therefore unfeasible to compare the performance of different time step functions for a single chaotic triple system. Instead, we generate an ensemble of 1000 random, equal-mass triple systems, drawn from a Plummer distribution \citep{Plummer_1911}. We integrate these triples for 300 N-body time units \citep{Heggie1986} (about 100 dynamical times, {where a dynamical time is the average time for a star to cross the system), after which about half of the ensemble has dissolved into a permanent binary/single configuration \citep[e.g.][]{TB_SPZ_2015}. By systematically varying the time step parameter, $\eta$, we obtain ensembles of solutions with varying accuracy. The integrator is fixed to be MCL4.

In Fig.~\ref{fig:triple}, we plot the results of the performance tests. First, we compare the robustness of the symmetrised time steps. We define the fraction $f_{\rm{stable}}$ as the fraction of simulations for which $h$ remains in the interval $\left[ T/4, 4T \right]$ throughout the simulation. We observe that the largest differences between the curves originate from the symmetrisation method, and that the different weighting methods produce approximately consistent results. Symmetrisation methods A1, B1 and C1 (all based on the linear function), produce the worst results. Methods A2 and B2 become increasingly robust for smaller time step parameters (towards larger CPU times). When comparing the symmetrisation families A, B and C, we find that family C (red) tends to lay above its counterparts in families A and B. Perfect stability is obtained for the methods based on the harmonic function.

We consider a time step criterion better than another one, if for the same CPU running time, it produces a better level of energy conservation. In Fig.~\ref{fig:triple}, we observe that stable symmetrised time steps indeed lead to better energy conservation. When comparing the harmonic methods (A3, B3 and C3), we find that they are competitive in performance. There is a slight advantage for A3, due to the fact that the calculation of the derivative of $T$ is not required. We observe that the level of energy conservation stalls around $10^{-8}$. This is probably due to the level of numerical precision used for these computations (double-precision) and the spread of the resulting round-off errors.\footnote{Round-off errors can be reduced by using regularisation methods \citep[e.g.][]{Mikkola_Tanikawa_1999} or by adopting arbitrary-precision arithmetic \citep[e.g.][]{TB_SPZ_2015}}.

\begin{figure*}
\centering
\begin{tabular}{c}
\includegraphics[height=0.4\textwidth,width=0.96\textwidth]{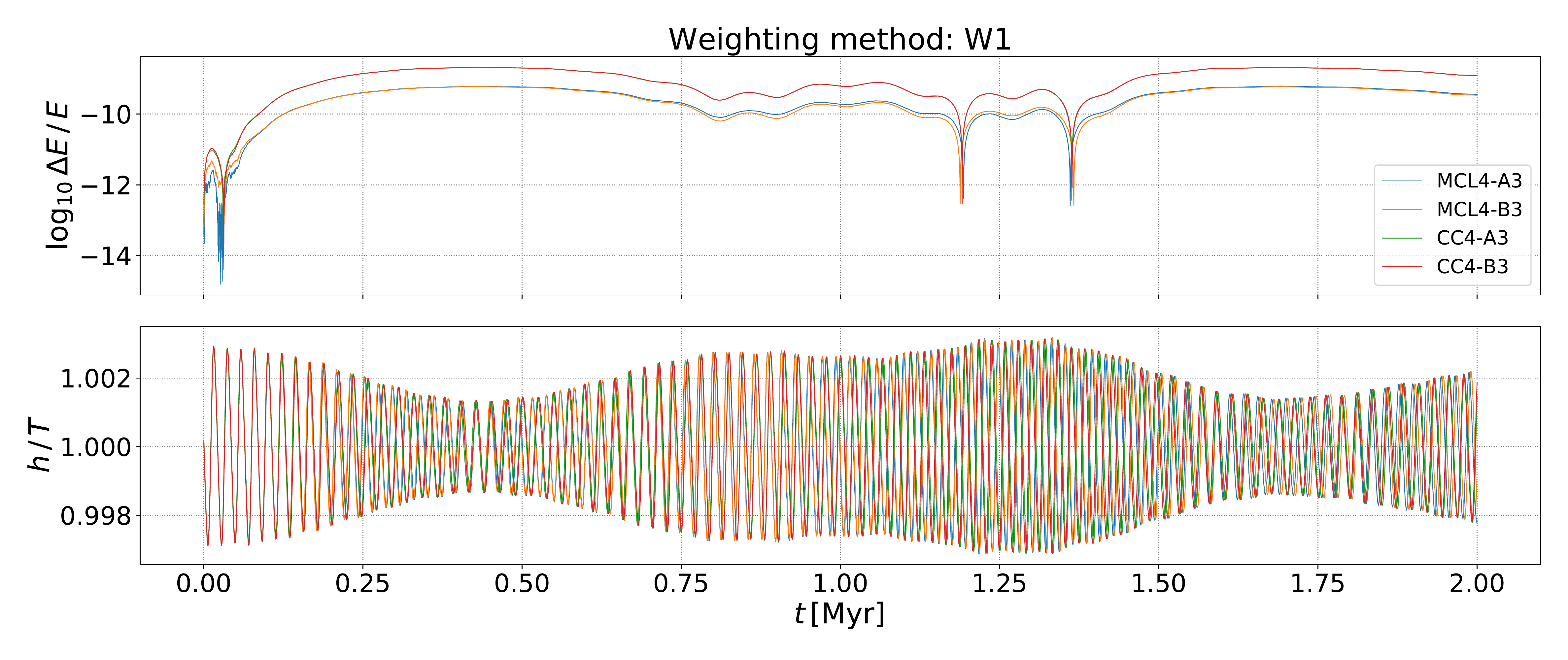} \\
\includegraphics[height=0.4\textwidth,width=0.96\textwidth]{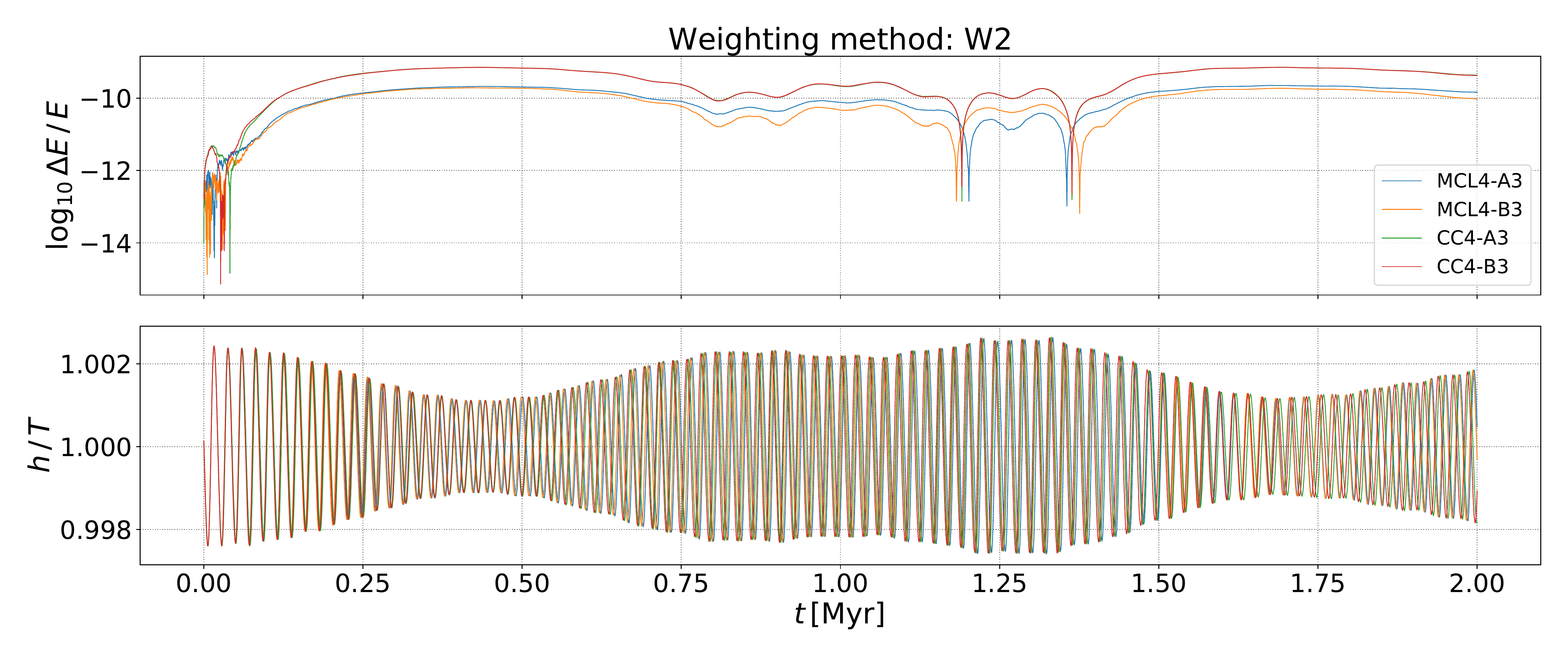} \\
\includegraphics[height=0.4\textwidth,width=0.96\textwidth]{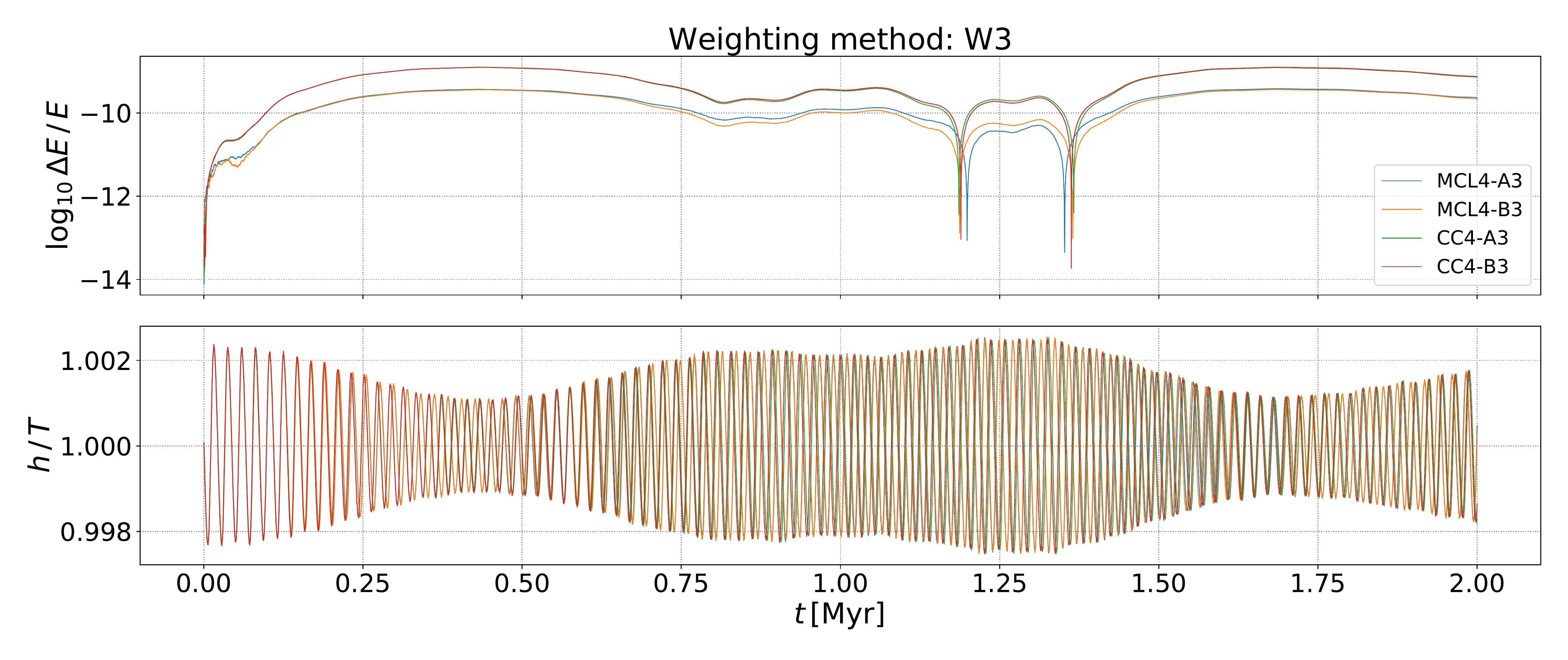} \\
\end{tabular}  
\caption{ We evolve the Solar System (Sun, 8 planets and Pluto) for two million years. We measure the time evolution of the ratio between the symmetrised time step and the time step function, $h/T$, as well as the absolute value of the relative energy conservation. We plot results obtained by both the MCL4 and CC4 integrators, and we vary the time step weighting method and symmetrisation method. The time step parameter is fixed to $\eta=2^{-6}$. Each method produces stable results, and the deviations among the different curves are very small. }
\label{fig:ss}
\end{figure*}

\begin{figure*}
\centering
\begin{tabular}{c}
\includegraphics[height=0.735\textwidth,width=0.98\textwidth]{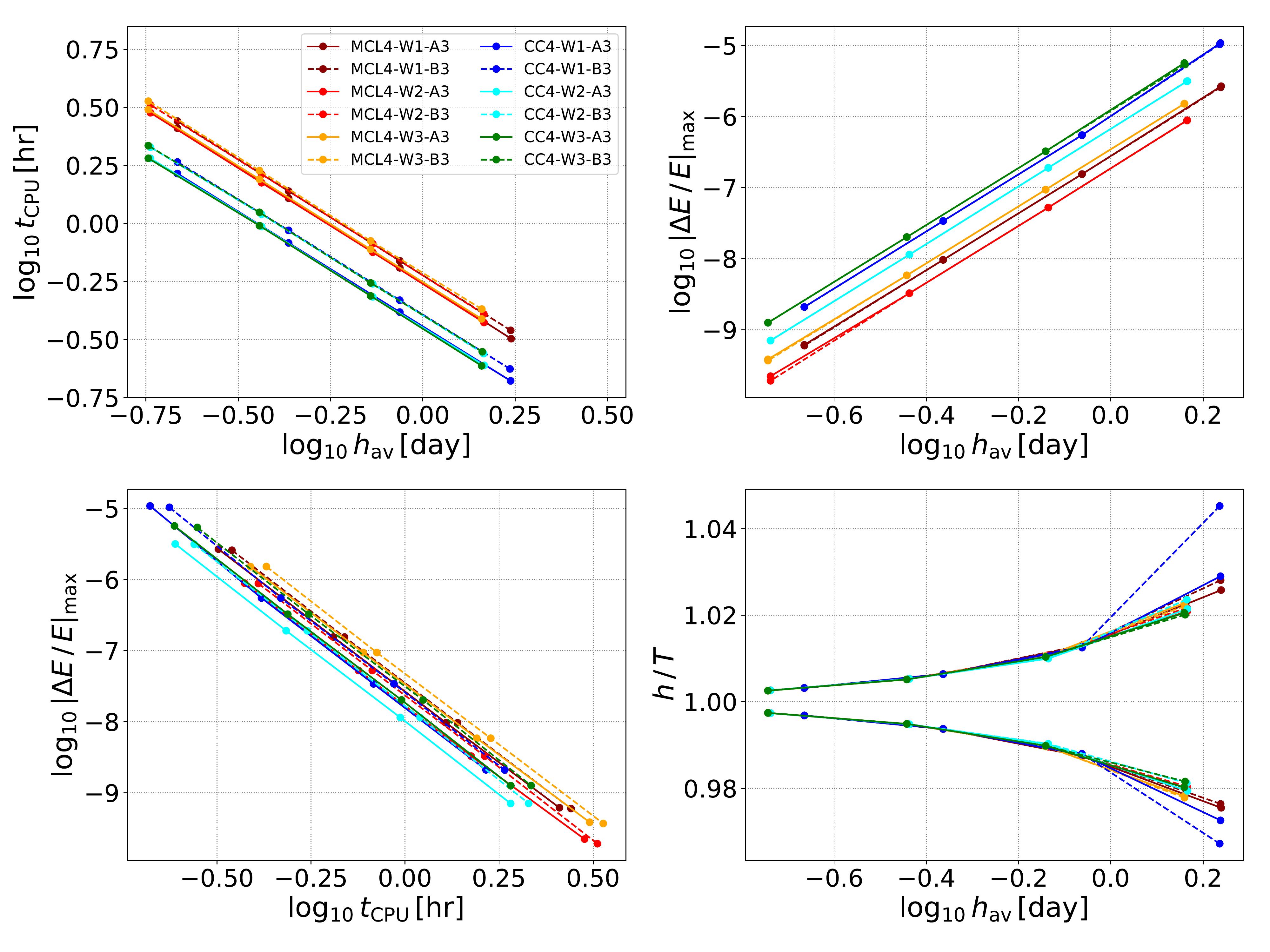} \\
\end{tabular}  
\caption{ Performance test for the Solar System. We plot the CPU running time (top left) and maximum, relative energy error (top right) as a function of average time step size. We also plot the maximum and minimum value of the ratio $h/T$ (bottom right). The performance comparison of energy error vs. CPU running time (bottom left) shows that the CC4-W2-A3 integration method performs the most optimal. }
\label{fig:ss_scaling}
\end{figure*}

\subsection{Nice model: unstable planetary systems}\label{sec:nice}

Symplectic integrators with a constant time step are often used for modelling planetary systems. This becomes problematic when the planetary system evolves towards an instability. This potentially results in planet-planet scattering, during which very close encounters are common. 
The transition from the planets orbiting the star, to close planet-planet deflections, is also a great test bed for adaptive time step functions.

As an example, we adopt an initial condition based on the Nice model, which assumes a compact, resonant chain for Jupiter, Saturn, Uranus and Neptune \citep[for more details see][]{Gomes_2005, Tsiganis_2005}. The initial semi-major axes of Jupiter is 6\,AU, Saturn is in a 3:2 resonance with Jupiter, Uranus in 3:2 with Saturn, and Neptune in 5:4 with Uranus (it is also possible to swap the initial position of Uranus and Neptune). We ignore the planetesimal disc, which would have damped the eccentricities and inclinations. This configuration is highly sensitive to small perturbations, and a slight change of the time step parameter, or changing the time step criterion, results in a different outcome.
We therefore consider an ensemble of 100 realisations, where the orbital angles are chosen randomly. The systems are evolved for a million years, at which point about $80\%$ of the systems have destabilised, resulting in at least one planet becoming unbound due to an ejection.

In Fig.~\ref{fig:nice}, we plot the result of the performance test. Focusing on the stability of the symmetrised time steps, we observe again that the symmetrisation methods based on the linear function are the most unstable. The only methods which are perfectly stable in this experiment are A3 and B3. The three different weighting methods produce roughly consistent results.

Focusing on the level of energy conservation, we observe that symmetrisation methods A2, B1 and C1 produce the largest outliers at large time steps. The other methods are competitive, but again, there is a slight edge for method A3 as the derivative of the time step does not have to be calculated.

\subsection{Solar System}\label{sec:ss}

In the previous two experiments, we found the consistent result that the symmetrisation methods A3 and B3 produce the best results. Here, we will compare these two methods further, while also varying the integration method (MCL4 and CC4).
We adopt a realisation of the Solar System \citep[from][]{Ito_2002}, including the Sun, the eight planets and Pluto ($N=10$). The Solar System is evolved up to two million years using pure Newtonian dynamics. 

In Fig.~\ref{fig:ss}, we plot the time evolution of the energy error and the ratio $h/T$. We observe that all three weighting methods produce robust results. Furthermore, the symmetrisation methods A3 and B3 produce consistent results when switching the integrator from MCL4 to CC4.

The statistical performance test is presented in Fig.~\ref{fig:ss_scaling}. 
Here, we define the average time step size, $h_{\rm{av}} = T/N_{\rm{step}}$, i.e. the simulation time divided by the number of integration steps.
We observe that with decreasing time steps, the ratio $h/T$ approaches unity for each of the solutions shown. We also observe that the integrator CC4 is faster than MCL4 for a fixed average time step size, but also less precise. In the unbiased comparison of energy error vs. CPU running time, we observe a slight advantage for the CC4 integrator. Secondly, we find a slightly better performance for the $W2$ weighting method, as well as for the A3 symmetrisation method. The most efficient combination is given by CC4-W2-A3.

\section{Conclusion}\label{sec:conclusion}

Symplectic integrators with adaptive time steps are appropriate for $N$-body systems with close encounter episodes. The secular drift of the energy error is reduced by combining the integration method with the use of a symmetrised time step, which depends symmetrically on the old and new states.
It then becomes important to have a time step, which behaves smoothly as a function of time. 
Our results confirm previous findings by \citet{Hands19} that a weighted average of all pairwise time steps, rather than adopting the minimal pairwise value, has a stabilising effect on the energy error, but only when combined with an explicit symmetrisation method. The manual addition of bounds to the symmetrised time step can serve as a safety net. The risk of an instability in the symmetrised time step is greatly reduced by implementing a robust method. Our experiments show that the methods based on the harmonic function, in particular A3 and B3, prove to be the most robust. Our Solar System experiment also showed a slightly improved efficiency for the $W2$ method. In the other two experiments the differences were more subtle, hinting towards the fact that as long as there is some implementation for smoothing, that symmetrisation methods will greatly benefit from it.

Too much smoothing however, reduces the dynamic range of the adaptive time step, and can result in weighted time steps that are too large to resolve close encounters. If we define the weight of a pairwise time step as the inverse of the time step raised to the power $n$, then the parameter $n$ controls the distribution of weights among the time steps. The optimal value of $n$ scales with the number of bodies as $n \propto \log_{10}N$. For $N$-body systems up to a few hundred bodies, we confirm a value of $n=10$ is sufficient.

It remains an open problem how to model chaotic $N$-body systems, such as {dense stellar systems}, without a secular drift in the energy error. Even with a symplectic integrator and symmetrised time steps, energy errors tend to drift. Error cancellation depends on the time symmetry of the integration, but also on whether the orbits themselves are time symmetric, i.e. periodic. For chaotic systems, the latter is generally not the case, so that a previous made error will generally not be undone in the future. 

However, even if we were able to remove the secular drift in the energy error for chaotic $N$-body systems, it is not guaranteed that the solution is also accurate \citep[][Fig.~3]{TB_SPZ_2015}. Small perturbations, whether physical or numerical, grow exponentially, and saturate after only a few Lyapunov time scales \citep[e.g.][]{1964ApJ...140..250M, 1993ApJ...415..715G, Hut_2002, TB_2020, ph421}.   
A comparison to numerically converged solutions, as can be obtained with a code such as \texttt{Brutus} \citep{SPZ_TB_2014, TB_SPZ_2015, TB_AM_2021}, are required to determine the statistical validity of $N$-body simulations with or without drifts in the energy error.  

\vspace{5mm}

\subsection*{Acknowledgements}

We are grateful to the referee Walter Dehnen for very helpful comments about this work and contributing through enlightening discussions, suggesting new ideas (in particular on the symmetrisation methods) and improving the overall presentation. The simulations were run on the Hydra computing cluster in Oxford.
  This project was supported by funds from the European Research Council (ERC) under the European Union’s Horizon 2020 research and innovation program under grant agreement No 638435 (GalNUC). 
This work was also supported by
CFisUC (UIDB/04564/2020 and UIDP/04564/2020), 
GRAVITY (PTDC/FIS-AST/7002/2020),
PHOBOS (POCI-01-0145-FEDER-029932), and
ENGAGE SKA (POCI-01-0145-FEDER-022217),
funded by COMPETE 2020 and FCT, Portugal.

%

\vspace{5mm}


\section*{Data Availability} 

The data underlying this article will be shared on reasonable request to the corresponding author.


\bibliography{agn}{}
\bibliographystyle{aasjournal}


\appendix


\section{Initial value for the previous symmetrised time step}\label{app:hprev}

Explicit time step symmetrisation methods keep track of the previous symmetrised time step, $h_{\rm{prev}}$, as discussed in Sec.~\ref{sec:methods}. An initial value has to be set in order to start the simulation. The simplest solution is to set $h_{\rm{prev}} = T$ initially. However, this turns out not to be the best option, and may affect the potential future divergence of the symmetrised time step from the time step function. A better alternative is to make use of one of the symmetrisation methods. If we consider a function, $F$, of the time step, we can estimate $h_{\rm{prev}}$ as

\begin{equation}
    F\left( h_{\rm{prev}} \right) = F\left( T \right) - \frac{1}{2} T \frac{dF}{dt}.
\end{equation}    
    
\noindent If we then apply the harmonic function, $F\left( h \right) = 1/h$, we obtain

\begin{equation}
    \frac{1}{h_{\rm{prev}}} = \frac{1}{T} + \frac{1}{2}\frac{\dot{T}}{T} = \frac{1}{T} \left( 1 + \frac{1}{2}\dot{T} \right).
\end{equation}
    
\noindent Or taking the inverse, we obtain:

\begin{equation}
    h_{\rm{prev}} = \frac{T}{1 + \frac{1}{2}\dot{T}}. 
\end{equation}

\noindent Note that this expression is similar to the approximate, symmetrised time step presented by \citet{Pelupessy_Janes_2012}, but rather in the negative time direction. In Fig.~\ref{fig:init}, we compare the two start-up values for an eccentric binary system. We observe that when $h_{\rm{prev}} = T$ and when we start away from pericenter or apocenter, that the symmetrised time step oscillates between two limits, but is biased with respect to the time step function. If we use the improved start-up value however, we confirm the bias is removed. We implement this improved initial value throughout our simulations. Other estimates for $h_{\rm{prev,0}}$ can be obtained using alternative functions and/or symmetrisation methods, or by replacing the term $T dF/dt$ by $h_{\rm{prev,0}} dF/dt$. However, since this only concerns the initialisation of $h_{\rm{prev}}$, and our improved estimate presented above works well, we expect the effect of an alternative initial value to be marginal.

\begin{figure*}
\centering
\begin{tabular}{c}
\includegraphics[height=0.33\textwidth,width=0.99\textwidth]{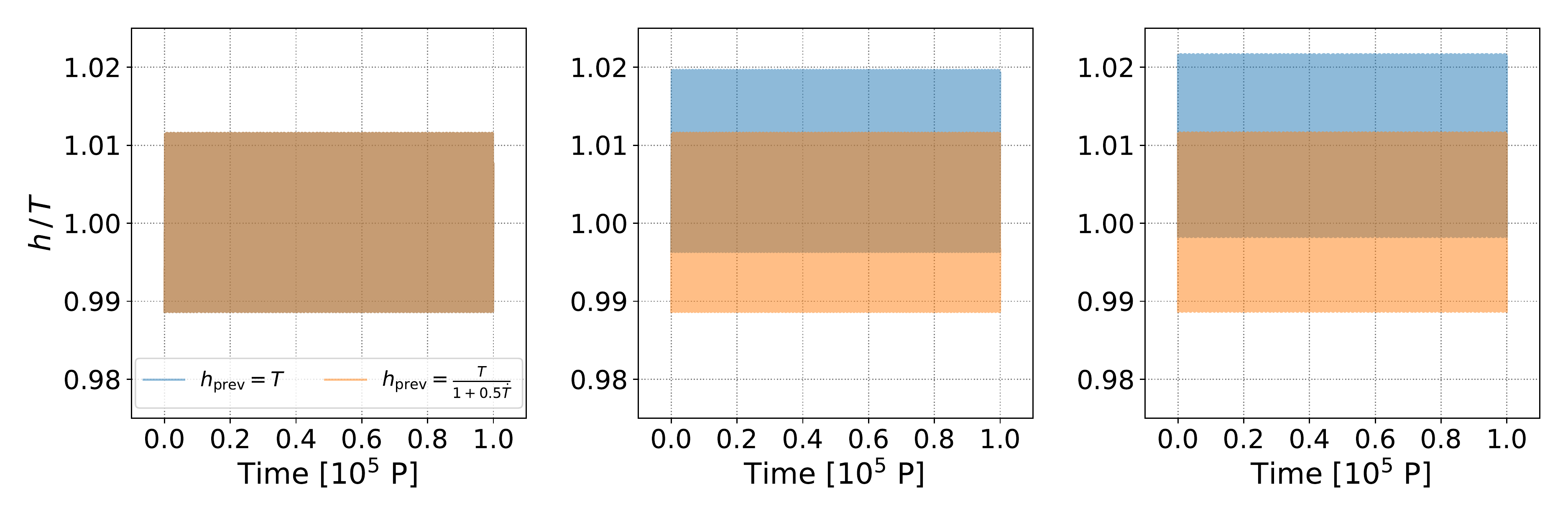} \\
\end{tabular}  
\caption{ We evolve an equal-mass and eccentric ($e=0.99$) binary system for $10^5$ orbital periods ($P$). We vary the initial true anomaly per panel: 0 (left), 90 (middle) and 175 (right) degrees. The symmetrisation method used is B3 (see Tab.~\ref{tab:dt}). We plot the time evolution of the ratio between the symmetrised time step, $h$, and the time step function, $T$. We compare two values for the initial previous symmetrised time step, $h_{\rm{prev}}$ (see legend). The criterion based on $T$ and $\dot{T}$ gives robust results in the sense that $h$ oscillates closely around $T$ for all three initial true anomalies. }
\label{fig:init}
\end{figure*}

\section{Smoothing with number of bodies}\label{app:logN}

Smoothing of the time step function has a positive effect on the functionality of time step symmetrisation methods. However, too much smoothing leads to time steps which are too large. Since the weighting methods sum over all pairs of bodies (for methods W1 and W2), or over each body individually (method W3), we expect a dependence between the amount of smoothing required and the number of bodies. In Sec.~\ref{sec:weights}, we derive that the minimum weight parameter should scale as $n \propto \log_{10}N$. The normalisation however, depends on the N-body configuration through the distribution of pairwise time step values. 

In order to test the scaling, we consider equal-mass Plummer spheres \citep{Plummer_1911} with $N$ ranging from 4 to 4,096. Each system is evolved for 1 N-body time unit, and for a specific value of $n$. By varying $n$ systematically, we measure how this changes the maximum deviation of the global time step, $T$, from the minimum pairwise time step, $T_{\rm{min}} = \min\left( T_{ij} \right)$. By interpolation, we calculate that value of $n$ for which the maximum deviation equals a factor $T / T_{\rm{min}} = 2$. In Fig.~\ref{fig:a_vs_N}, we plot this critical value of $n$ as a function of $N$ for the $W3$ weighting method. We confirm that the scaling approximately follows the expectation. Furthermore, for a particle number up to a few thousand, we confirm that $n > 6$ is required. The choice of $n=10$ adopted here, and also by \citet{Hands19}, thus leads to a sufficient amount of smoothing to facilitate the symmetrisation process, while avoiding too much smoothing.

\begin{figure}
\centering
\begin{tabular}{c}
\includegraphics[height=0.42\textwidth,width=0.48\textwidth]{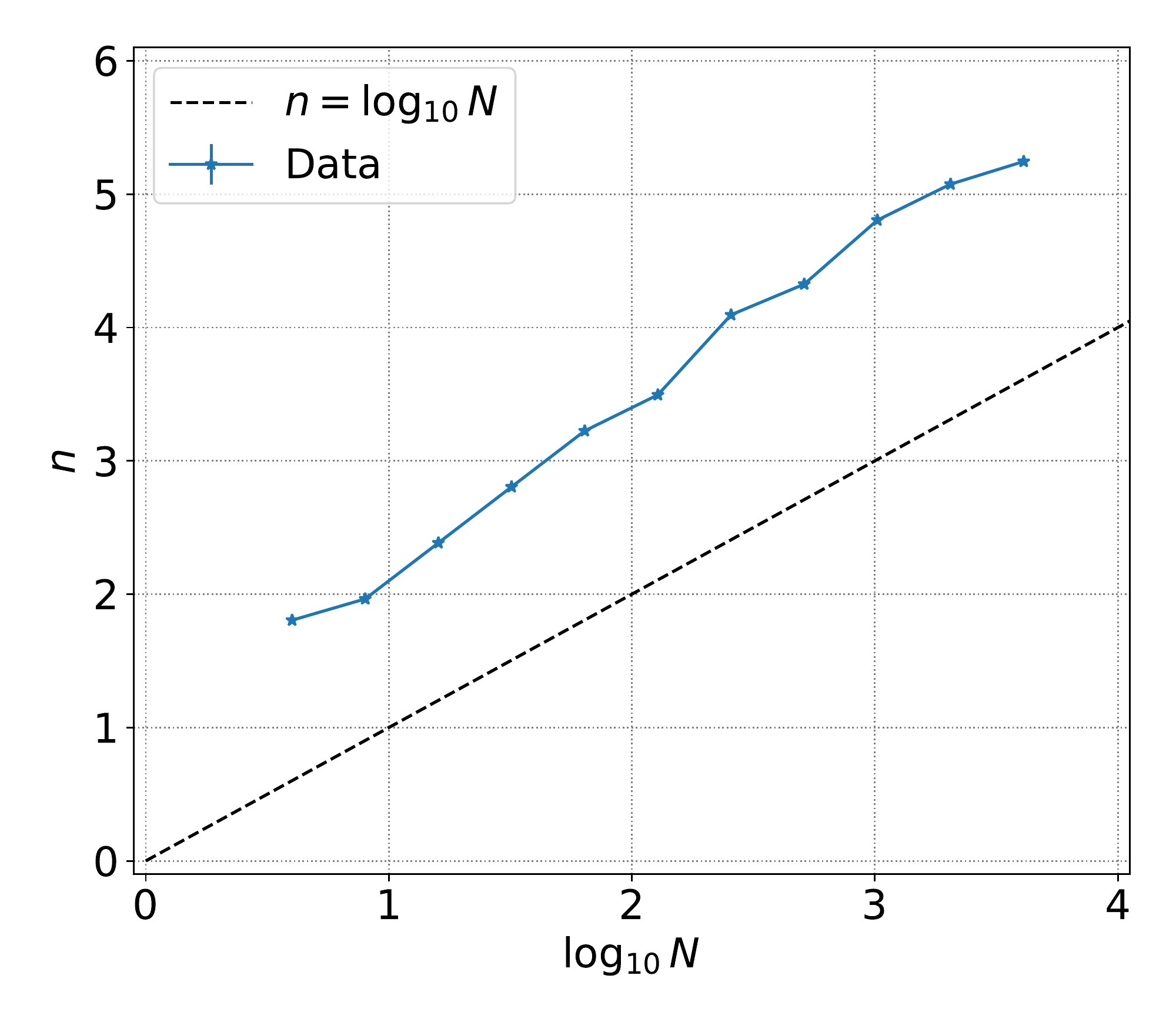} \\
\end{tabular}  
\caption{ We consider equal-mass Plummer spheres consisting of $N$ bodies. For each value of $N$ we measure the value of the weight parameter, $n$, for which the global time step, $T$, is equal to twice the minimum pairwise time step, i.e. $T\left( n \right) = 2 \min\left\{ T_{\rm{ij}} \right\}$, using the $W3$ method. We confirm the expectation that approximately $n \sim \log_{10}\,N$. The normalisation however, depends on the N-body configuration through the distribution of time steps. }
\label{fig:a_vs_N}
\end{figure}



\end{document}